\documentstyle[12pt]{article}

\def\hybrid{\topmargin -20pt	\oddsidemargin 0pt
	\headheight 0pt	\headsep 0pt
	\textwidth 6.25in	
	\textheight 9.5in	
	\marginparwidth .875in
	\parskip 5pt plus 1pt	\jot = 1.5ex}

\hybrid

\def\baselinestretch{1.2}

\catcode`\@=11

\def\marginnote#1{}
\newcount\hour
\newcount\minute
\newtoks\amorpm
\hour=\time\divide\hour by60
\minute=\time{\multiply\hour by60 \global\advance\minute by-\hour}
\edef\standardtime{{\ifnum\hour<12 \global\amorpm={am}%
	\else\global\amorpm={pm}\advance\hour by-12 \fi
	\ifnum\hour=0 \hour=12 \fi
	\number\hour:\ifnum\minute<10 0\fi\number\minute\the\amorpm}}
\edef\militarytime{\number\hour:\ifnum\minute<10 0\fi\number\minute}

\def\draftlabel#1{{\@bsphack\if@filesw {\let\thepage\relax
   \xdef\@gtempa{\write\@auxout{\string
      \newlabel{#1}{{\@currentlabel}{\thepage}}}}}\@gtempa
   \if@nobreak \ifvmode\nobreak\fi\fi\fi\@esphack}
	\gdef\@eqnlabel{#1}}
\def\@eqnlabel{}
\def\@vacuum{}
\def\draftmarginnote#1{\marginpar{\raggedright\scriptsize\tt#1}}

\def\draft{\oddsidemargin -.5truein
	\def\@oddfoot{\sl preliminary draft \hfil
	\rm\thepage\hfil\sl\today\quad\militarytime}
	\let\@evenfoot\@oddfoot	\overfullrule 3pt
	\let\label=\draftlabel
	\let\marginnote=\draftmarginnote
   \def\@eqnnum{(\theequation)\rlap{\kern\marginparsep\tt\@eqnlabel}%
\global\let\@eqnlabel\@vacuum}  }

\def\preprint{\twocolumn\sloppy\flushbottom\parindent 2em
	\leftmargini 2em\leftmarginv .5em\leftmarginvi .5em
	\oddsidemargin -.5in	\evensidemargin -.5in
	\columnsep .4in	\footheight 0pt
	\textwidth 10.in	\topmargin  -.4in
	\headheight 12pt \topskip .4in
	\textheight 6.9in \footskip 0pt
	\def\@oddhead{\thepage\hfil\addtocounter{page}{1}\thepage}
	\let\@evenhead\@oddhead	\def\@oddfoot{}	\def\@evenfoot{} }

\def\numberbysection{\@addtoreset{equation}{section}
	\def\theequation{\thesection.\arabic{equation}}}

\def\underline#1{\relax\ifmmode\@@underline#1\else
	$\@@underline{\hbox{#1}}$\relax\fi}

\def\titlepage{\@restonecolfalse\if@twocolumn\@restonecoltrue\onecolumn
     \else \newpage \fi \thispagestyle{empty}\c@page\z@
	\def\thefootnote{\fnsymbol{footnote}} }

\def\endtitlepage{\if@restonecol\twocolumn \else \newpage \fi
	\def\thefootnote{\arabic{footnote}}
	\setcounter{footnote}{0}}  

\catcode`@=12
\relax

\def\figcap{\section*{Figure Captions\markboth
	{FIGURECAPTIONS}{FIGURECAPTIONS}}\list
	{Figure \arabic{enumi}:\hfill}{\settowidth\labelwidth{Figure 999:}
	\leftmargin\labelwidth
	\advance\leftmargin\labelsep\usecounter{enumi}}}
 \relax
\def\tablecap{\section*{Table Captions\markboth
	{TABLECAPTIONS}{TABLECAPTIONS}}\list
	{Table \arabic{enumi}:\hfill}{\settowidth\labelwidth{Table 999:}
	\leftmargin\labelwidth
	\advance\leftmargin\labelsep\usecounter{enumi}}}
 \relax
\def\reflist{\section*{References\markboth
	{REFLIST}{REFLIST}}\list
	{[\arabic{enumi}]\hfill}{\settowidth\labelwidth{[999]}
	\leftmargin\labelwidth
	\advance\leftmargin\labelsep\usecounter{enumi}}}
 \relax
\makeatletter
\newcounter{pubctr}
\def\publist{\@ifnextchar[{\@publist}{\@@publist}}
\def\@publist[#1]{\list
	{[\arabic{pubctr}]\hfill}{\settowidth\labelwidth{[999]}
	\leftmargin\labelwidth
	\advance\leftmargin\labelsep
	\@nmbrlisttrue\def\@listctr{pubctr}
	\setcounter{pubctr}{#1}\addtocounter{pubctr}{-1}}}
\def\@@publist{\list
	{[\arabic{pubctr}]\hfill}{\settowidth\labelwidth{[999]}
	\leftmargin\labelwidth
	\advance\leftmargin\labelsep
	\@nmbrlisttrue\def\@listctr{pubctr}}}
 \relax
\makeatother
\newskip\humongous \humongous=0pt plus 1000pt minus 1000pt

\newif\ifdtup

\relax

\def\thefootnote{\fnsymbol{footnote}}
\def\be{\begin{equation}}
\def\ee{\end{equation}}
\def\ba{\begin{eqnarray}}
\def\ea{\end{eqnarray}}

\begin{document}
\renewcommand{\theequation}{\thesection.\arabic{equation}}
\newcommand{\beq}{\begin{equation}}
\newcommand{\eeq}[1]{\label{#1}\end{equation}}
\newcommand{\ber}{\begin{eqnarray}}
\newcommand{\eer}[1]{\label{#1}\end{eqnarray}}
\begin{titlepage}
\begin{center}

\hfill CERN-TH/96-121\\
\hfill hep-th/9605043\\

\vskip .5in

{\large \bf SOLITONS OF AXION--DILATON GRAVITY}

\vskip 0.6in

{\bf Ioannis Bakas}
\footnote{e-mail addresses: BAKAS@SURYA11.CERN.CH, BAKAS@NXTH04.CERN.CH}\\
\vskip .1in

{\em Theory Division, CERN, 1211 Geneva 23, Switzerland\\
and\\
Department of Physics, University of Patras, 26110 Patras, Greece}
\footnote{Permanent address}\\
\vskip .6in

\end{center}

\vskip .6in

\begin{center} {\bf ABSTRACT } \end{center}
\begin{quotation}\noindent
We use soliton techniques of the two-dimensional reduced
$\beta$-function equations to obtain non-trivial string
backgrounds from flat space. These solutions are characterized
by two integers $(n, m)$ referring to the soliton numbers of
the metric and axion-dilaton sectors respectively. We show that
the Nappi-Witten universe associated with the
$SL(2) \times SU(2) / SO(1, 1) \times U(1)$
CFT coset arises as an $(1, 1)$
soliton in this fashion for certain values of the moduli
parameters, while for other values of the soliton
moduli we arrive at the
$SL(2)/SO(1, 1) \times SO(1, 1)^2$ background.
Ordinary 4-dim black-holes
arise as 2-dim $(2, 0)$ solitons, while the Euclidean
worm-hole background is described as a $(0, 2)$ soliton
on flat space. The soliton transformations correspond
to specific elements of the string Geroch group. These
could be used as starting point for exploring the role of
U-dualities in string compactifications to two dimensions.
\end{quotation}
\vskip.3cm
CERN-TH/96-121 \\
May 1996\\
\end{titlepage}
\vfill
\eject

\def\baselinestretch{1.2}
\baselineskip 16 pt

\section{Introduction}
\noindent
Duality symmetries in string theory arise as discrete
remnants of continuous groups of transformations of the
lowest order effective theory. These symmetries have
received a lot of attention, as they can also provide
non-perturbative information about string theory. The
most common examples are T and S-dualities, but it has
also become clear recently that U-dualities can be
sucessfully used to explore various generalized
equivalences among superstrings [1, 2].

Dimensional reduction offers the possibility to intertwine
the T and S moduli, and hence construct large groups of
solution generating symmetries in three and two dimensions.
For example, the reduction from 4 to 3 dimensions gives rise
to an $O(2, 2)$ group [3], while the reduction from 4 to 2
dimensions leads to an infinite dimensional group of the
lowest order effective theory, the current group $\hat{O} (2, 2)$
[4].
These results can be regarded as straightforward generalization
of similar structures found by Geroch in the space of
solutions of vacuum Einstein equations with one or two
commuting isometries [5], but now they also include apart
from the metric $G_{\mu \nu}$ the antisymmetric tensor field
$B_{\mu \nu}$ and the dilaton $\Phi$. The coset space structure
of the scalar fields in various dimensionally reduced
supergravity theories was known before (see for instance [6, 7, 8]
and references therein). More recent
is the realization that T and S dualities are embedded in the
corresponding continuous hidden symmetry groups.
In a heterotic string context it means that the reduction
from 10 to 3 dimensions gives rise to an $O(8, 24)$ group [7, 9],
while the reduction from 10 to 2 dimensions leads to
$\hat{O} (8, 24)$ [10, 11]. It is then natural to expect that
the 2-dim sector of string theory will be quite rich in symmetry,
having as U-duality an appropriately chosen discrete subgroup
of the underlying string Geroch current group. Up to this day,
however, very little progress has been made in this particular
direction, since proving the conjectured U-dualities and
understanding their action on the full spectrum of superstring
models based on these effectively 2-dim backgrounds turns into a
difficult problem.

In this paper we consider string models with target space
$M_{4} \times K$, where $M_{4}$ is a 4-dim spacetime with
signature $-+++$ and $K$ is some internal space, which is
usually represented by a conformal field theory (CFT), so that
the total central charge is critical. We focus on cosmological
backgrounds $M_{4}$ with non-trivial $G_{\mu \nu}$, $B_{\mu \nu}$
and $\Phi$ that arise as solutions of the lowest order effective
theory,
\be
S_{eff} = \int_{M_{4}} d^4 X \sqrt{- \det G} \left( R -
2 ({\nabla}_{\mu} \Phi)^2 - {1 \over 12} e^{-4 \Phi}
H_{\mu \nu \rho}^2 \right) .
\ee
Here the theory is defined directly in the Einstein frame,
which is related to the $\sigma$-model frame by
$G_{\mu \nu}^{(\sigma)} = e^{2 \Phi} G_{\mu \nu}$, and the
effective cosmological constant is taken zero.
It will be convenient for later use to trade $B_{\mu \nu}$ with
the axion field $b$, which is consistently defined in the
Einstein frame as follows:
\be
{\partial}_{\mu} b = {1 \over 6} e^{-4 \Phi} \sqrt{- \det G} ~
{{\epsilon}_{\mu}}^{\nu \rho \sigma} H_{\nu \rho \sigma} ~ .
\ee
$H_{\mu \nu \rho}$ is the field strength of $B_{\mu \nu}$ and
${\epsilon}_{0123} = 1$. In $M_{4}$ with signature$-+++$ we may
further define the complex conjugate fields
$S_{\pm} = b \pm i e^{-2 \Phi}$, which provide the natural variables
of
the S-moduli. Later we will also consider string backgrounds with
Euclidean signature $++++$.

There is a very limited number of exact CFT backgrounds, which
to lowest order in ${\alpha}^{\prime}$ provide solutions of (1.1).
The most characteristic examples of this type are the two pairs of
WZW coset
models
\be
{SL(2) \times SU(2) \over SO(1, 1) \times U(1)} ~ , ~~~~~
{SL(2) \over SO(1, 1)} \times SO(1, 1)^2 ~ ,
\ee
\be
{SL(2) \over SO(1, 1)} \times {SU(2) \over U(1)} ~ , ~~~~~
SL(2) \times SO(1, 1) ~.
\ee
The first model in (1.3) depends on a free parameter that defines the
gauging of the coset, and it is particularly interesting in
string cosmology as it describes a closed inhomogeneous
expanding and recollapsing universe [12] (see also [13] for
some earlier ideas). The other three models are the Lorentzian
counterparts obtained by analytic continuation of the $D=4$,
$\hat{c} = 4$, $N = 4$ superconformal backgrounds
$C^{(4)} = SU(2)/U(1) \times U(1)^2$,
${\Delta}^{(4)} = SL(2)/U(1) \times SU(2)/U(1)$ and
$W^{(4)} = SU(2) \times U(1)$ (the throat of a worm-hole)
respectively
with appropriately
chosen background charges [14, 15, 16]. All these models exhibit
two commuting Killing isometries. It has been established with
the aid of $O(2, 2)$
transformations that
the first model in (1.3) is related to the first model in (1.4) [17],
and similarly the other two models of the series
are T-dual to each other [15, 18].

Our contribution is to connect the two gravitational backgrounds
associated with the Nappi-Witten universe
$SL(2) \times SU(2) / SO(1, 1) \times U(1)$ and
$SL(2)/SO(1, 1) \times SO(1, 1)^2$ to the trivial
flat space background $F^{(4)}$ with zero $B_{\mu \nu}$ and
$\Phi$, by considering a specially chosen six dimensional
moduli space of backgrounds within the entire set of solutions
of (1.1) with two commuting Killing isometries. This is
technically achieved by performing first the 2-dim reduction of
the effective theory (1.1), and then employing solitonic
constructions that are available for the resulting integrable
system of equations (both for the metric and the axion-dilaton
sectors). As it turns out, the simplest $(1, 1)$ soliton
configuration on $F^{(4)}$, or more precisely its T-dual face,
will be sufficient to describe the semiclassical backgrounds
of these two coset models as 2-dim solitons for appropriate
choices of the six moduli parameters. The solitonic dressing
of (the dual of) $F^{(4)}$ in this paper is analogous
to the solitonic dressing of Kasner type metrics that
were studied by Belinski and Sakharov in the context of general
relativity many years ago [19]. In the context of pure gravity
these authors gave a very interesting derivation of 4-dim
black holes as 2-dim double soliton solutions on flat space. Further
work has also appeared in the literature, which describes the
physically very interesting situation of two
colliding gravitational plane waves
in terms of 2-dim solitons [20]. Given the wide
applicability of these methods, it is also natural
to consider the explicit form of solitons in axion-dilaton
gravity and attempt a reinterpretation of known solutions, in
particular those that correspond to exact CFT backgrounds,
in this context.

It is interesting to note that the soliton dressing of a
given configuration corresponds to a specific choice of finite
group element of the Geroch group (see for instance [21]), and
hence in the string context we have found the way to generate
the two exact CFT backgrounds (1.3) from flat space by U-duality
(viewed as a continuous group of transformations at this point).
In a sense one may then say that the dualities
create a universe, the Nappi-Witten cosmological solution in
this case. We will also consider Euclidean string backgrounds
and show for example that the worm-hole solution, which is an axionic
instanton of the 10-dim heterotic theory, arises as a 2-dim $(0, 2)$
soliton on flat space.
Certainly, many more connections can be made between different string
backgrounds using the inverse scattering method of the 2-dim
reduced sector, and we hope to return to them in the near future.

In section 2 we briefly discuss the dimensionally reduced string
background equations and outline the construction of soliton
solutions using the integrability of the resulting 2-dim
$\sigma$-models. In section 3 we construct the most general
$(1, 1)$ soliton solution on a T-dual face of Minkowski space
and determine the choice of moduli parameters that correspond to
the CFT backgrounds $SL(2) \times SU(2)/ SO(1, 1) \times U(1)$
and $SL(2)/SO(1, 1) \times SO(1, 1)^2$. In section 4 we describe
the ordinary 4-dim black-holes as 2-dim $(2, 0)$ solitons on
flat space, where the soliton moduli correspond to the mass,
rotation and NUT parameters of the most general stationary
axisymmetric solution. In section 5 the Euclidean worm-hole
background is interpreted as a 2-dim $(0, 2)$ soliton in the same
context. Section 6 contains our conclusions and some directions
for further work on the subject. We argue that the present
results could be most importantly used as starting point for
exploring the role of U-dualities in compactifications of
string theory to two dimensions.

\section{The reduced theory and its solitons}
\setcounter{equation}{0}
\noindent
The effective theory (1.1) describes the coupling of an ordinary
$SL(2)/U(1)$ $\sigma$-model to 4-dim gravity, which is manifest in
the
axion-dilaton formulation using the field variables
$S_{\pm} = b \pm i e^{-2 \Phi}$. This axion-dilaton $\sigma$-model
has Lorentzian signature if $M_{4}$ is Euclidean, but in the case
of interest here, where $M_4$ is Lorentzian, the $SL(2)/U(1)$
$\sigma$-model is Euclidean. Hence, it is convenient to parametrize
the axion-dilaton sector by the symmetric matrix
\be
\lambda = e^{2 \Phi} \left(\begin{array}{ccc}
1 & & b \\
  & &   \\
b & & b^2 + e^{-4 \Phi} \end{array} \right)
\ee
so that $\det \lambda = 1$.

Following [4] (and references therein) we consider gravitational
string
backgrounds of cosmological type with two commuting Killing
isometries,
so that the target space metric is restricted by the ansatz
\be
ds^2 = f(X^0 , X^1 ) \left( - (dX^0)^2 + (dX^1)^2 \right) +
g_{AB} (X^0 , X^1 ) dX^A dX^B ~; ~~~~ A, ~ B= 2, ~3 ,
\ee
and also $b(X^0 , X^1 )$, $\Phi (X^0 , X^1 )$. For notational
convenience
we introduce light-cone coordinates
\be
\eta = {1 \over 2} (X^0 - X^1 ) ~, ~~~~~ \xi = {1 \over 2} (X^0 + X^1
) ~.
\ee
Then, using the defining relations (1.2) this ansatz amounts to
choosing
special backgrounds with only $B_{23} \neq 0$, in which case the
axion
equations simplify to
\be
{\partial}_{\xi} b = {e^{-4 \Phi} \over \sqrt{\det g}}
{\partial}_{\xi}
B_{23} ~, ~~~~~
{\partial}_{\eta} b = - {e^{-4 \Phi} \over \sqrt{\det g}}
{\partial}_{\eta}
B_{23} ~.
\ee
This class of backgrounds will be quite sufficient for the present
purposes
of our work.

The reduced string background equations that follow from (1.1) read
as follows
in the Einstein frame:
\be
{\partial}_{\eta} (\sqrt{\det g} ~ g^{-1} {\partial}_{\xi} g) +
{\partial}_{\xi} (\sqrt{\det g} ~ g^{-1} {\partial}_{\eta} g) = 0 ~,
\ee
\be
{\partial}_{\eta} (\sqrt{\det g} ~ {\lambda}^{-1} {\partial}_{\xi}
\lambda ) +
{\partial}_{\xi} (\sqrt{\det g} ~ {\lambda}^{-1} {\partial}_{\eta}
\lambda ) = 0~.
\ee
They are two essentially decoupled $SL(2)/U(1)$ 2-dim $\sigma$-models
of Ernst
type (both having Euclidean signature), one for the metric sector $g$
and
one for the axion-dilaton $\lambda$. Since $\sqrt{\det g}$ satisfies
the 2-dim
wave equation ${\partial}_{\eta} {\partial}_{\xi} \sqrt{\det g} = 0$,
we may choose without loss of generality
\be
X^0 = \sqrt{\det g} \equiv \alpha ~, ~~~~~ X^1 = \beta
\ee
for the corresponding pair of its conjugate solutions. From now on we
assume the
special choice of coordinates (2.7), using $\alpha = \sqrt{\det g}$
and $\beta$
instead of $X^0$ and $X^1$ in (2.2), and reserve the notation $X^0$,
$X^1$ for
more general coordinate systems. Sometimes we will also denote the
remaining two
coordinates by $z$ and $w$ instead of $X^2$ and $X^3$, in order to
make more
uniform our presentation in the following sections.

We recall that the differential equations for the conformal factor
$f$ are linear
of first order,
\be
{\partial}_{\xi} (\log f) = -{1 \over \alpha} + {\alpha \over 4} {\rm
Tr} \left(
(g^{-1} {\partial}_{\xi} g)^2 + ({\lambda}^{-1} {\partial}_{\xi}
\lambda )^2
\right) ,
\ee
\be
{\partial}_{\eta} (\log f) = -{1 \over \alpha} + {\alpha \over 4}
{\rm Tr} \left(
(g^{-1} {\partial}_{\eta} g)^2 + ({\lambda}^{-1} {\partial}_{\eta}
\lambda )^2
\right)
\ee
and so, once a solution $(g, \lambda )$ of the two Ernst
$\sigma$-models is known,
$f$ can be simply determined integrating by quadratures.
All these calculations are performed in the Einstein frame, where the
decoupling
of (2.5) and (2.6) takes place, but the results can be easily
translated in the
$\sigma$-model frame of string theory.

The non-linear $\sigma$-models of the 2-dim reduced theory are known
to be
integrable, and it is precisely this property that is responsible for
having an
infinite dimensional symmetry group, the (string) Geroch group,
acting on the
space of classical solutions. Because of integrability the 2-dim
$\sigma$-models
admit soliton solutions, which can be constructed explicitly on any
given background that acts as a seed for the solitons. On any string
background
we may actually construct a whole series of solitonic excitations $(n
, m)$, where
$n$ and $m$ denote the soliton numbers of the $g$ and $\lambda$
sectors
respectively.
Here we briefly review the essential ingredients of the soliton
technique for
the Ernst $\sigma$-model using only the metric sector of the theory,
but the
construction is exactly the same for the axion-dilaton sector since
$\sqrt{\det g} ~ \lambda$ satisfies the same equation (2.6) as
$\lambda$.
The new solutions that arise in this fashion are solitons in the
2-dim sense,
and although 4-dim string backgrounds can be reconstructed from the
$(n , m)$
data, the resulting configurations are not necessarily solitons of
the
4-dim world saturating the Bogomol'ny bound, and thus they are
generically
quantum mechanically unstable.
Next, we revisit the soliton framework of Belinski-Sakharov
[19], because their technique is not widely known to string
theorists.

Consider the following linear system of $(2 \times 2)$-matrix
differential
equations:
\be
D_{1} \Psi = {A \over l- \alpha} \Psi ~, ~~~~~
D_{2} \Psi = {B \over l+ \alpha} \Psi ~,
\ee
where $\Psi (\eta , \xi ;l)$ is a complex matrix function depending
on a spectral
parameter $l$ that takes values in the whole complex plane, and
\be
A = - \alpha {\partial}_{\xi} g g^{-1} , ~~~~~
B = \alpha {\partial}_{\eta} g g^{-1} .
\ee
Also, the differential operators are
\be
D_1 = {\partial}_{\xi} - 2 {l \over l - \alpha} {\partial}_{l} ~ ,
{}~~~~~
D_2 = {\partial}_{\eta} + 2 {l \over l + \alpha} {\partial}_{l}
\ee
and clearly they commute, $[D_{1} ~ , ~ D_{2}] = 0$. The system
(2.10) is
compatible provided that $g$ satisfies the Ernst equation (2.5) for
$\sqrt{\det g} = \alpha$. We actually assume
\be
\Psi (\eta , \xi ; l = 0) = g(\eta , \xi ) ~,
\ee
and so $\Psi$ can be regarded as
a suitable generalization of $g$ with spectral parameter.

Let $g_{0} (\eta , \xi )$ be a known solution and let $\Psi_{0} (\eta
, \xi ; l)$
be the corresponding solution of the linear system (2.10). If we
assume that
other solutions $g$ exist such that
\be
\Psi (l) = \chi (l) {\Psi}_{0} (l) ~,
\ee
then $\chi (\eta , \xi ; l)$ has to satisfy the system of equations
\be
D_1 \chi = {1 \over l - \alpha} (A \chi - \chi A_0 ) , ~~~~~
D_2 \chi = {1 \over l + \alpha} (B \chi - \chi B_0 ) ,
\ee
where $A_0$, $B_0$ are the currents (2.11) of a seed metric $g_0$. If
we
manage to find an appropriate $\chi (l)$, then according to (2.13)
and (2.14),
a new solution will be obtained
\be
g(\eta , \xi) = \chi (l = 0) g_0 ~.
\ee
There are a few technical assumptions on $\chi$, namely the reality
condition
on the real $l$-line, $\bar{\chi} (\bar{l}) = \chi (l)$, and $\chi
(\infty) = 1$.

The $n$-soliton excitations of a given seed background $g_0$ are very
special in that
$\chi$ has a simple pole structure in the complex $l$-plane
\be
\chi (\eta , \xi ; l) = 1 + \sum_{k=1}^{n}
{R_{k} (\eta , \xi ) \over
l - {\mu}_k (\eta , \xi ) } ~ .
\ee
The residue and pole functions can be determined substituting (2.17)
in (2.15)
and start comparing the pole structure on the left and right hand
sides.
The details are rather lengthy and we skip them here. We only give
the final
result that will be used later for explicit computations. The poles
are roots
of the algebraic equation
\be
{\mu}_{k}^2 + 2(\beta - C_{0}^{(k)}) {\mu}_{k} + {\alpha}^2 = 0 ~,
\ee
where $C_{0}^{(k)}$ are arbitrary numerical constants (moduli),
and ${\mu}_k$ satisfy the differential equations in $\eta$, $\xi$,
\be
{\partial}_{\eta} {\mu}_k = {2 {\mu}_k \over \alpha + {\mu}_k } ~ ,
{}~~~~~
{\partial}_{\xi} {\mu}_k = {2 {\mu}_k \over \alpha - {\mu}_k } ~.
\ee

The residue matrices $R_k$ are degenerate having the component form
\be
{(R_k)}_{AB} = N_A^{(k)} M_B^{(k)} ~,
\ee
where the 2-component vector $M^{(k)}$ is given using the inverse of
$\Psi_0$
at $l = {\mu}_k$,
\be
M_B^{(k)} = \sum_{A} C_A^{(k)} {\Psi}_{0}^{-1} (\eta , \xi ; l =
{\mu}_k )_{AB} ~.
\ee
$C^{(k)}$ is an arbitrary
constant 2-component vector, which together with $C_{0}^{(k)}$
provide all the moduli parameters of the general solitonic excitation
of $g_0$.
The other 2-component vectors $N^{(k)}$ have more complicated form
and can not
be explicitly written with the same ease. $N^{(k)}$ are the solution
vectors of
the $n$-th order linear system of algebraic equations
\be
\sum_{l=1}^{n} {\Gamma}_{kl} N_{A}^{(l)} = {1 \over {\mu}_k }
\sum_{B} M_B^{(k)}
{(g_{0})}_{AB} ~,
\ee
where the $n \times n$ matrix $\Gamma$ was determined by Belinski and
Sakharov
\be
{\Gamma}_{kl} = {1 \over {\mu}_k {\mu}_l - {\alpha}^2} \sum_{A, B}
M_{A}^{(k)}{(g_0)}_{AB} M_{B}^{(l)} ~.
\ee

Therefore, putting it all together we arrive at a concrete expression
for the $n$-soliton excitation of $g_0$, namely
\be
g(\eta , \xi) = \left(1 - \sum_{k=1}^{n} {R_k \over {\mu}_k }
\right) g_0 ~.
\ee
A final issue is the overall normalization of the dressing matrix
$\chi$.
Using this last equation we find
$\det g = {\alpha}^{2n +2}{\mu}_1^{-2} {\mu}_2^{-2} \cdots
{\mu}_n^{-2}$, which differs from $\det g_0 = {\alpha}^2$. Agreement
is achieved
by scaling $\chi$ with ${\mu}_1 {\mu}_2 \cdots {\mu}_n / {\alpha}^n$,
and this is
what we will assume from now on.
The properly normalized $n$-soliton dressing matrices of $g_0$,
\be
\chi (l = 0) = {{\mu}_1 {\mu}_2 \cdots {\mu}_n \over {\alpha}^n }
\left(1 - \sum_{k=1}^{n} {R_k \over {\mu}_k } \right) ,
\ee
define specific group elements of the Geroch group. The normalization
(2.25) is
introduced to achieve consistency with the standard formulation of
Geroch
transformations that preserve $\det g$.

Summarizing, if we apply this proceduce to any given string
background $g_{0}$, ${\lambda}_{0}$
(more precisely $\alpha {\lambda}_0$ to be in exact analogy for both
sectors),
we will obtain a generic $(n , m)$ solitonic excitation with $3n +
3m$ continuous
moduli. One of the difficulties to implement this construction in
practice,
apart from the problem of inverting the
corresponding matrices $\Gamma$ for large $n$ and $m$, is to find
the explicit solution ${\Psi}_0$ of (2.10) (and its
$\lambda$-counterpart)
when an arbitrary background is used as seed. For this reason we will
start from
very simple seed solutions, knowing ${\Psi}_0$,
and use the soliton technique to construct (and hence reinterpret
in this context) the more
complicated solutions that exist in the literature.

The Geroch group of the metric sector is
the loop group $\hat{SL} (2)$, and when both sectors are taken into
account
the Geroch group becomes $\hat{SL} (2) \times \hat{SL} (2) = \hat{O}
(2, 2)$.
It is known in this case how to obtain the entire algebra by
successive
intertwining of continuous T and S transformations [4]. We briefly
mention here
that 4-dim backgrounds with two commuting isometries exhibit the
obvious
$O(2, 2)$ group of transformations on the space of solutions. These
transformations
are non-locally realized in the axion-dilaton formulation of the
theory and
their generators are embedded in the algebra of the string Geroch
group
$\hat{O} (2, 2)$ as follows: we use the zero mode subalgebra of the
$g$-$\hat{SL} (2)$,
say $T_{+}^0$, $T_{-}^0$, $T_{0}^0$, and the non-locally realized
$SL(2)$
subalgebra of the $\lambda$-$\hat{SL} (2)$, $\tilde{T}_{+}^{-1}$,
$\tilde{T}_{0}^0$, $\tilde{T}_{-}^{1}$ that includes the
$\pm 1$ modes. The continous analogue of
the S-duality $SL(2)$
transformations are locally realized in the axion-dilaton formulation
and correspond to the zero mode generators $\tilde{T}_{+}^0$,
$\tilde{T}_{-}^0$, $\tilde{T}_{0}^0$ of the $\lambda$-$\hat{SL} (2)$.
Hence, by
intertwining $O(2, 2)$ with S we can generate after an infinite
number of steps
the entire $\hat{SL} (2)$ algebra of the axion-dilaton sector. To
generate the
other $\hat{SL} (2)$
we interchange the field variables $g \leftrightarrow \alpha \lambda$
and perform the same intertwining procedure. The exchange of the two
sectors is
a legitimate operation in this case
because both $\sigma$-models have Euclidean
signature,
and this is also a $Z_2$ symmetry of the 2-dim
reduced string background equations,
leaving $f$ unaffected.

The $n$-soliton matrices (2.25) could also be described in terms of
specific elements of the infinite dimensional group of Geroch
transformations obtained by other approaches [21].
We will not attempt here to decompose them in terms of more
fundamental
operations associated to successive intertwining of T and S
transformations,
but we note as an important property their commutativity in the
following
sense: an $(n + n^{\prime})$-soliton can either be constructed
directly
from a seed background or it can be viewed as an
$n^{\prime}$-solitonic
excitation of the $n$-soliton, and similarly for $n \leftrightarrow
n^{\prime}$.
Since there is a
systematic understanding of the group elements of soliton dressing,
we think
that is worth exploring further the precise meaning and the
consequences of U-duality in this particular
sector of string theory. Of course much work remains to be done in
this direction.

\section{(1, 1) solitons and CFT backgrounds}
\setcounter{equation}{0}
\noindent
According to the general framework of the previous section we
may compute the simplest 1-soliton solution of the Ernst
$\sigma$-model, say (2.5), using as seed metric
\be
g_{0} = \left(\begin{array}{ccc}
        {\alpha}^{2s_{1}} & & 0 \\
        0 & & {\alpha}^{2s_{2}} \end{array} \right) ;
{}~~~~~ s_{1} + s_{2} = 1 ~.
\ee
In a purely gravitational context this choice of the seed metric
corresponds to a Kasner cosmological background. There are two
special cases in this family, namely
\be
s_{1} = 0 , ~~ s_{2} = 1 ; ~~~~~ s_{1} = s_{2} = {1 \over 2} ~,
\ee
which correspond to flat space (in polar coordinates) and an
isotropic
universe respectively. Using (2.10) we may determine ${\Psi}_0$ for
this
background,
\be
{\Psi}_0 (l) = \left( \begin{array}{ccc}
(l^2 + 2 \beta l + {\alpha}^2)^{s_{1}} & & 0 \\
  &  &  \\
  &  &  \\
0 &  & (l^2 + 2 \beta l + {\alpha}^2)^{s_{2}} \end{array} \right) .
\ee
We see clearly that ${\Psi}_0 (l = 0) = g_{0}$ as required on general
grounds
(2.13).

We will first derive the general form of the 1-soliton solution, and
then
make various specializations according to the connections we would
like
to make later with 4-dim CFT backgrounds.
The 1-soliton background on $g_{0}$ (3.1) is obtained using only one
pole located at
\be
\mu = {\mu}_{\pm} = C_{0} - \beta \pm \sqrt{(C_{0} - \beta)^2 -
{\alpha}^2} ~.
\ee
This pole is positioned on the real axis of the complex $l$-plane
provided
that
\be
(C_{0} - \beta)^2 \geq {\alpha}^2 ~.
\ee
Otherwise we will be forced to consider more complicated double
soliton
solutions, since complex poles always come in pairs.
Then, the physical 1-soliton matrix (after normalization with $\mu /
\alpha$)
reads
\be
g_{1} = {\alpha \over \mu A} \left(\begin{array}{ccc}
C_{2}^2 (2C_{0} \mu)^{2 s_{1}} {\mu}^2 + C_{1}^2 (2C_{0} \mu)^{2
s_{2}}
{\alpha}^{4s_{1}} & & 2C_{0} C_{1} C_{2} \mu ({\alpha}^2 - {\mu}^2)
\\
  & & \\
  & & \\
2 C_{0} C_{1} C_{2} \mu ({\alpha}^2 - {\mu}^2) & & C_{1}^2 (2C_{0}
\mu)^{2s_{2}}
{\mu}^2 + C_{2}^2 (2 C_{0} \mu)^{2s_{1}} {\alpha}^{4s_{2}}
\end{array} \right) ,
\ee
where
\be
A = C_{1}^2 (2C_{0} \mu)^{2s_{2}} {\alpha}^{2s_{1}} + C_{2}^2
(2 C_{0} \mu)^{2 s_{1}} {\alpha}^{2 s_{2}} ~.
\ee

If we were to apply the same construction to the axion-dilaton Ernst
$\sigma$-model
(2.6), we should have scaled $\alpha$ in front of the seed matrix
(3.1), since
$g_{0}$ behaves the same way as $\sqrt{\det g} ~ {\lambda}_{0} =
\alpha {\lambda}_{0}$.
Let us begin with a background having
\be
e^{-2 {\Phi}_{0}} = {\alpha}^{2s_{2}^{\prime} - 1} ~ ; ~~~~~ b_{0} =
0
\ee
with $s_1^{\prime} + s_2^{\prime} = 1$ as well.
Scaling out $\alpha$ from the general form of the 1-soliton solution,
we
find that the new axion-dilaton system is given by the configuration
\ba
e^{-2 {\Phi}_{1}} & = & \mu {{C_{1}^{\prime}}^2 (2C_{0}^{\prime}
{\mu}^{\prime})^{2s_{2}^{\prime}} {\alpha}^{2s_{1}^{\prime}} +
{C_{2}^{\prime}}^2 (2C_{0}^{\prime}
{\mu}^{\prime})^{2s_{1}^{\prime}} {\alpha}^{2s_{2}^{\prime}} \over
{C_{1}^{\prime}}^2 (2C_{0}^{\prime}
{\mu}^{\prime})^{2s_{2}^{\prime}} {\alpha}^{4s_{1}^{\prime}} +
{C_{2}^{\prime}}^2 (2C_{0}^{\prime}
{\mu}^{\prime})^{2s_{1}^{\prime}} {{\mu}^{\prime}}^2} ~,\\
b_{1} & = & {2C_{0}^{\prime} C_{1}^{\prime} C_{2}^{\prime}
{\mu}^{\prime}
({\alpha}^2 - {{\mu}^{\prime}}^2) \over
{C_{1}^{\prime}}^2 (2C_{0}^{\prime}
{\mu}^{\prime})^{2s_{2}^{\prime}} {\alpha}^{4s_{1}^{\prime}} +
{C_{2}^{\prime}}^2 (2C_{0}^{\prime}
{\mu}^{\prime})^{2s_{1}^{\prime}} {{\mu}^{\prime}}^2} ~,
\ea
where the primes are used to distinguish the parameters of the
axion-dilaton system
from those of the metric moduli.

Suppose now we are combining both sectors to construct the $(1, 1)$
soliton starting
from the following solution of the string background equations in the
Einstein frame:
\ba
ds^2 & = & -d{\alpha}^2 + d{\beta}^2 + \alpha (dz^2 + dw^2 ) ~,\\
b_{0} & = & 0 ~, ~~~~~ e^{-2 {\Phi}_{0}} = \alpha ~.
\ea
This particular choice of the seed background is very special in that
is
T-dual to $F^{(4)}$, i.e., the flat space metric with zero dilaton
and
antisymmetric tensor fields. To see this we translate (3.11) in the
$\sigma$-model
frame and perform T-duality with respect to the Killing coordinate
$\beta$, which
yields the purely gravitational background
\be
ds^2 = -{1 \over \alpha} d {\alpha}^2 + \alpha d {\beta}^2 + dz^2 +
dw^2 ~.
\ee
Introducing coordinates
\be
x = 2 \sqrt{\alpha} ~ \cosh {\beta \over 2} ~ , ~~~~~
y = 2 \sqrt{\alpha} ~ \sinh {\beta \over 2} ~ ,
\ee
the metric (3.13) assumes the flat space form
$ds^2 = -dx^2 + dy^2 + dz^2 + dw^2$. Actually it is immediately
recognized that
(3.14) is a Rindler transformation of the 2-dim Minkowski space $(x,
y)$ with
$(\log \alpha , ~ \beta)$ providing the corresponding pair of Rindler
coordinates.
Hence, we start from the 2-dim Rindler wedge times a flat 2-dim
Euclidean
space parametrized by the other two coordinates $(z, w)$, and use its
T-dual
face as seed string background.

We notice that our ansatz (3.11), (3.12) for the seed background
imply the following
choice of the Kasner type parameters for the two sectors,
\ba
g &:& s_{1} = s_{2} = {1 \over 2} ~, \\
\lambda &:& s_{1}^{\prime} = 0 ~, ~~~~ s_{2}^{\prime} = 1 ~.
\ea
Then the resulting $(1, 1)$ soliton simplifies considerably
and in the Einstein frame is given by
\ba
g_{1} &=& {1 \over C_{1}^2 + C_{2}^2} \left(\begin{array}{ccc}
C_{1}^2 {{\alpha}^2 \over \mu} + C_{2}^2 \mu & & C_{1} C_{2}
\left({{\alpha}^2 \over \mu} - \mu \right) \\
  &  &  \\
C_{1} C_{2} \left({{\alpha}^2 \over \mu} - \mu \right) & & C_{1}^2
\mu
+ C_{2}^2 {{\alpha}^2 \over \mu} \end{array} \right) ,\\
e^{-2 {\Phi}_{1}} &=& {1 \over 4{C_{0}^{\prime}}^2 {C_{1}^{\prime}}^2
+
{C_{2}^{\prime}}^2} \left(4 {C_{0}^{\prime}}^2 {C_{1}^{\prime}}^2
{\mu}^{\prime}
+ {C_{2}^{\prime}}^2 {{\alpha}^2 \over {\mu}^{\prime}} \right) ,\\
b_{1} &=& {2C_{0}^{\prime} C_{1}^{\prime} C_{2}^{\prime} \over
4{C_{0}^{\prime}}^2 {C_{1}^{\prime}}^2 + {C_{2}^{\prime}}^2}
\left({{\alpha}^2 \over
{\mu}^{\prime}} - {\mu}^{\prime} \right) .
\ea
As for the conformal factor $f$, which follows by integration of
(2.8), (2.9), we find
after some lengthy computation the
result
\be
f_{1} = {\mu (4{C_{0}^{\prime}}^2 {C_{1}^{\prime}}^2
{{\mu}^{\prime}}^2 +
{C_{2}^{\prime}}^2 {\alpha}^2 ) \over ({\alpha}^2 - {\mu}^2 )
({\alpha}^2 -
{{\mu}^{\prime}}^2 )}
\ee
up to an overall numerical factor, whereas $f_{0} = 1$ by inspecting
(3.11).

There is an ambiguity to choose ${\mu}_{+}$ or ${\mu}_{-}$ in
(3.17)--(3.19), but since
\be
{\mu}_{\pm} = {{\alpha}^2 \over {\mu}_{\mp}} ~,
\ee
the two choices yield the same result provided that in the metric
soliton moduli space
$(C_{1} , C_{2}) \rightarrow (-C_{2} , C_{1})$. Similarly in the
axion-dilaton
sector the two choices ${\mu}_{\pm}^{\prime}$ are equivalent provided
that
$(C_{1}^{\prime} , C_{2}^{\prime}) \rightarrow
(-C_{2}^{\prime}/2C_{0}^{\prime} , ~
2 C_{0}^{\prime} C_{1}^{\prime})$. Hence in the following we may
choose without loss of
generality
\be
\mu = C_{0} - \beta + \sqrt{(C_{0} - \beta)^2 - {\alpha}^2} ~ , ~~~~
{\mu}^{\prime} = C_{0}^{\prime} - \beta + \sqrt{(C_{0}^{\prime} -
\beta)^2 - {\alpha}^2} ~.
\ee

Next, we show how to obtain the string backgrounds associated to the
two coset models
$SL(2) \times SU(2) / SO(1, 1) \times U(1)$ and $SL(2)/SO(1, 1)
\times SO(1, 1)^2$
by making appropriate choices of the moduli parameters in the 6-dim
space of solutions
we have obtained.

\noindent
(i) \underline{Nappi-Witten universe} :

In the $g$-sector of the general $(1, 1)$ soliton solution we choose
\be
C_{1} = 0 ~,
\ee
which gives rise to a diagonal metric with components
\be
g_{zz} = C_{0} - \beta + \sqrt{(C_{0} - \beta)^2 - {\alpha}^2} ~ ,
{}~~~~
g_{ww} = C_{0} - \beta - \sqrt{(C_{0} - \beta)^2 - {\alpha}^2}
\ee
independent of $C_{2}$. For the axion-dilaton sector we set
\be
C_{0}^{\prime} = 1 ~, ~~~~~ {C_{2}^{\prime} \over 2C_{1}^{\prime}} =
{\sin \theta -1 \over \cos \theta} ~,
\ee
where $\theta$ is an arbitrary numerical constant. Hence, choosing
${\mu}_{+}^{\prime}$
we also fix
\ba
e^{-2 \Phi} & = & 1 - \beta + \sin \theta \sqrt{(1 - \beta)^2 -
{\alpha}^2} ~,\\
b & = & \cos \theta \sqrt{(1 - \beta)^2 - {\alpha}^2} ~,
\ea
while the conformal factor in the Einstein frame is determined
according to (3.20).

We claim that this solution corresponds to the cosmological
background found by
Nappi and Witten while considering the $SL(2) \times SU(2)/SO(1, 1)
\times U(1)$ CFT coset.
In this regard, the numerical parameter $\theta$
that was introduced in (3.25) will be shown to describe
the arbitrariness in the gauging of this coset. For this purpose we
also choose
\be
C_{0} = -1 ~,
\ee
thus describing the same $\theta$-dependent string background for any
point in the
soliton moduli space that is restricted by (3.23), (3.25) and (3.28).

The construction is rather formal up to now, while making various
seemingly unjustified
choices of the free parameters. At this point we introduce
coordinates $X^0$, $X^1$ in
terms of $\alpha$, $\beta$ given by
\be
\alpha = \sin 2X^0 \sin 2X^1 ~, ~~~~~ \beta = \cos 2X^0 \cos 2X^1 ~,
\ee
thus also restricting the range of $\alpha$ and $\beta$ as $X^0$ and
$X^1$ range from $0$ to $\pi / 2$.
This is a good choice because
\be
e^{-2 \Phi} = 1 - \cos 2X^0 \cos 2X^1 + \sin \theta (\cos 2X^0 - \cos
2X^1 )
\ee
is manifestly real and positive, as should be expected for an honest
dilaton field.
For the axion we find
\be
b = \cos \theta (\cos 2X^0 - \cos 2X^1 ) ~.
\ee
At first sight it seems that these choices are not good for the
metric sector (3.24),
since
\be
g_{zz} = -4 {\sin}^2 X^0 {\sin}^2 X^1 ~, ~~~~~ g_{ww} = -4 {\cos}^2
X^0 {\cos}^2 X^1 ~,
\ee
and the signature turns out to be $--$ instead of $++$. Recall,
however, the way
we have obtained the physical metric in the soliton construction of
section 2. There,
we had to scale $\chi (l=0)$ accordingly so that $\det g = \det g_{0}
= {\alpha}^2$.
The scaling was $\mu/\alpha$ for the 1-soliton, but equally well we
could have taken
$- \mu/\alpha$. The latter choice renders the signature of $g$
physical, i.e. $++$, and
there is no contradiction.

To make exact contact with the Nappi-Witten cosmological background
we introduce
coordinates $X^2$ and $X^3$ by scaling
\be
w = \sqrt{{1 + \sin \theta \over 2}} ~ X^2 ~, ~~~~~ z = \sqrt{{1 -
\sin \theta \over 2}}
{}~ X^3
\ee
and compute the full metric in the $\sigma$-model frame. The final
result reads
\ba
ds_{(\sigma)}^2 & = & -(dX^0)^2 + (dX^1)^2 + {2 \over 1 - \cos 2X^0
\cos 2X^1 +
\sin \theta (\cos 2X^0 - \cos 2X^1 )} \cdot \nonumber\\
 & & \left((1 + \sin \theta) {\cos}^2 X^0 {\cos}^2 X^1 (dX^2)^2 +
(1 - \sin \theta) {\sin}^2 X^0 {\sin}^2 X^1 (dX^3)^2 \right) .
\ea
We also compute the antisymmetric tensor field from the axion and
find that all its
components are zero apart from
\be
B_{23} = {1 \over 2} ~ {\cos 2X^0 - \cos 2X^1 + \sin \theta (1 - \cos
2X^0 \cos 2X^1 )
\over 1 - \cos 2X^0 \cos 2X^1 + \sin \theta (\cos 2X^0 - \cos 2X^1 )}
{}~.
\ee
This is precisely the result that was obtained in the semi-classical
limit of the
$SL(2) \times SU(2) / SO(1, 1) \times U(1)$ coset model
having an arbitrary parameter $\theta$
that specifies the gauging [12].

So, according to this, the Nappi-Witten universe can be created from
flat space
starting from (a suitably restricted part of) 
the Rindler wedge, performing a T-duality
transformation and then an
$(1, 1)$ soliton dressing. Consequently, our procedure completely
determines the
group element of the string Geroch group
$\hat{O} (2, 2)$ that connects classically the two
backgrounds. The Nappi-Witten background describes a closed expanding
and
recontracting universe as $X^0$ varies from $0$ (big bang) to $\pi/2$
(big crunch).
These two authors performed an in depth analysis of the model noting
that for
\be
{1 - \sin \theta \over \cos \theta} = {\rm rational ~ number}
\ee
$X^0=0$ or $\pi/2$ are orbifold singularities.
Also, away from the
special values $X^1=0$ or $\pi/2$ respectively, these are
singularities in the causal
structure of space-time rather than curvature singularities.

This cosmological solution is positioned in the entire moduli space
of $(1, 1)$
solitons as follows: consider the 3-dim subspace with axis labelled
by $C_2$, $C_1^{\prime}$
and $C_2^{\prime}$, while keeping the other coordinates fixed to
their chosen values
$C_1 = 0$, $C_0^{\prime} = 1 =-C_0$; if we draw all 2-dim planes
having the $C_2$-line
as common axis, then every point on each such plane will correspond
to the same solution,
while rotating planes change $\theta$. In this description the
criterion (3.35) for
having orbifold singularities is equivalent to considering rational
values for the slope of the
solution plane, which is given by $C_2^{\prime}/C_1^{\prime}$
according to (3.25).

Concluding we mention that the points of the moduli space
with the same restrictions as before, but
with
$C_2^{\prime} = 0$,
yield (a suitable analytic continuation of) the background
$SL(2)/SO(1, 1) \times SU(2)/U(1)$. Using the parametrization (3.29)
it
follows from our general expression that the axion field is zero,
the dilaton field is
\be
e^{-2 \Phi} = {\cos}^2 X^0 {\sin}^2 X^1 ~,
\ee
and the metric in the $\sigma$-model frame is diagonal,
\be
ds_{(\sigma)}^2 = -(dX^0)^2 + (dX^1)^2 + {\cot}^2 X^1 (dX^2)^2 +
{\tan}^2 X^0 (dX^3)^2 ~.
\ee
It describes a suitable real form of the direct product of two 2-dim
black-hole cosets. This background was used in [14] to obtain the
complete Nappi-Witten solution by $O(2, 2)$ transformations. If
$(1 - \sin \theta)/ \cos \theta$ takes only integer values, the
transformation is in $O(2, 2; Z)$ and the underlying backgrounds are
equivalent as exact conformal field theories. It is interesting to
note
that the transformation that provides the $(1, 1)$ soliton dressing
already contains in it the corresponding $O(2, 2)$ group elements;
but
it also contains much more that allow for a flat space derivation of
these
CFT backgrounds.

\noindent
(ii) \underline{The coset $SL(2)/SO(1, 1) \times SO(1, 1)^2$} :

Following the same construction as above we will now specify other
points in the moduli
space of soliton solutions (3.17)--(3.19) that lead to the
semi-classical geometry of
the $SL(2)/SO(1, 1) \times SO(1, 1)^2$ coset.

We choose $C_1 = 0$ for the metric sector,
thus arriving at the same expression (3.24) as before,
while for the axion-dilaton sector we let $C_2^{\prime} = 0$. In this
case we find
\ba
e^{-2 \Phi} &=& C_0^{\prime} - \beta + \sqrt{(C_{0}^{\prime} -
\beta)^2 -{\alpha}^2} ~,\\
b &=& 0 ~.
\ea
We furthermore let
\be
C_{0} = C_{0}^{\prime} ~,
\ee
and introduce the coordinate transformation
\be
\alpha = {1 \over 2} e^{2 X^1} \sinh 2X^0 ~ , ~~~~~ \beta = C_{0} -
{1 \over 2}
e^{2X^1} \cosh 2X^0 ~,
\ee
which clearly has $(C_{0} - \beta)^2 \geq {\alpha}^2$ as required for
reality.
We find in this parametrization
\be
e^{-2 \Phi} = e^{2X^1} {\cosh}^2 X^0 ~,
\ee
while the antisymmetric tensor field is zero and the metric is
diagonal. In the $\sigma$-model
frame, also setting $w = X^2$ and $z = X^3$, the metric assumes the
form
\be
ds_{(\sigma)}^2 = -(dX^0)^2 + (dX^1)^2 + {\tanh}^2 X^0 (dX^2)^2 +
(dX^3)^2 ~,
\ee
and the resulting background coincides with the geometry of the coset
$SL(2)/SO(1, 1) \times SO(1, 1)^2$ as it was advertized.

The worm-hole background will be discussed separately in section 5
using
2-dim solitons in Euclidean space.

\section{Black-holes as 2-dim (2, 0) solitons}
\setcounter{equation}{0}
\noindent
In this section we briefly review for completeness the interpretation
of ordinary 4-dim black-holes as 2-dim $(2, 0)$ solitons, filling up
some of the intermediate steps of the calculation [19] as well.
We use as starting point the flat space metric in polar coordinates,
\be
ds^2 = d{\alpha}^2 + d{\beta}^2 + {\alpha}^2 d {\varphi}^2 -
d{\tau}^2 ~,
\ee
for which the matrix $\Psi$ of the linearized system (2.10) for
stationary axisymmetric metrics is
\be
{\Psi}_0 (l) = \left(\begin{array}{ccc}
{\alpha}^2 - 2 \beta l -l^2 & & 0\\
                 &  &  \\
0 & & -1 \end{array} \right).
\ee

We point out a few differences between this case and the cosmological
setting
of the two previous sections. Here, the 2-dim space $(\alpha ,
\beta)$ has Euclidean
signature, while the $\sigma$-model $g$ is Lorentzian. As a result,
one has to take
into account various sign changes in order to adopt the general
soliton construction
to stationary axisymmetric metrics; in particular (2.18) changes to
\be
{\mu}_{k}^2 + 2 (\beta - C_{0}^{(k)}) {\mu}_k - {\alpha}^2 = 0 ~,
\ee
and the factor ${\mu}_k {\mu}_l - {\alpha}^2$ in (2.23) changes to
${\mu}_k {\mu}_l + {\alpha}^2$. Analogous changes have to be
introduced in the
linearized system of equations (2.10)--(2.12) and the differential
equations
(2.19) for the poles ${\mu}_k$. Also, the $n$-soliton transformation
of a
seed background $g_{0}$ yields $\det g = (-1)^n {\alpha}^{2n + 2}
{\mu}_1^{-2}
{\mu}_2^{-2} \cdots {\mu}_n^{-2}$ and the normalization (2.25) is the
same
as before for $n$ even. If $n$ is odd, however, the signature of the
soliton
metric $g$ changes sign to $++$, which is is not acceptable. For this
reason
the simplest physical soliton to construct is the double soliton
solution on
flat space (4.1).

The present version of the formalism will also become relevant in the
next section,
while considering Euclidean gravitational solutions of the string
background
equations. In that case the axion-dilaton system corresponds to a
Lorentzian
$\sigma$-model $\lambda$, and the explicit construction of its
solitons will
require the modifications we are considering here.

After some calculation we find that the general $(2, 0)$ soliton on
the purely
gravitational background (4.1) has metric components
\be
g_{\tau \varphi} = 2({\mu}_1 - {\mu}_2)({\alpha}^2 + {\mu}_1 {\mu}_2)
{N_1 \over D} ~, ~~~~~ g_{\tau \tau} = - 4 {\mu}_1 {\mu}_2 {N_2 \over
D} ~,
\ee
where
\ba
N_1 & = & A \Sigma ({\alpha}^2 + {\mu}_1^2)({\alpha}^2 + {\mu}_2^2) +
AM {\alpha}^2 ({\mu}_1^2 - {\mu}_2^2) - B \Sigma 
({\alpha}^4 - {\mu}_1^2
{\mu}_2^2) ~,\\
N_2 & = & {\Sigma}^2 ({\alpha}^2 + {\mu}_1 {\mu}_2)^2 - A^2
{\alpha}^2
({\mu}_1 - {\mu}_2)^2 ,\\
D & = & \left( (\Sigma + M) {\mu}_1 + (\Sigma - M) {\mu}_2 \right)^2
({\alpha}^2 + {\mu}_1 {\mu}_2)^2 + \nonumber\\
& & \left( (A-B) {\alpha}^2 - (A+B)
{\mu}_1 {\mu}_2 \right)^2 ({\mu}_1 - {\mu}_2)^2~ ,
\ea
while $g_{\varphi \varphi}$ is determined by the condition $\det g =
-{\alpha}^2$.
Also the corresponding conformal factor turns out to be
\be
f = {{\mu}_1 {\mu}_2 \over 4 ({\mu}_1 - {\mu}_2)^2 ({\alpha}^2 +
{\mu}_1^2)
({\alpha}^2 + {\mu}_2^2) ({\alpha}^2 + {\mu}_1 {\mu}_2)^2} D ~,
\ee
whereas $f_0 = 1$.
In the above expressions the parameters $A$, $B$, $M$ and $\Sigma$
are the special
combinations of the $C^{(k)}$ moduli,
\ba
C_1^{(1)} C_2^{(2)} & = & (\Sigma - M) C_0^{(1)}
~,
{}~~~~~~~
C_{1}^{(1)} C_1^{(2)} = 2 (A-B) C_0^{(1)} C_0^{(2)} ~, \nonumber\\
C_2^{(1)} C_1^{(2)} & = & -(\Sigma + M)
C_0^{(2)}~,
{}~~~~~
C_2^{(1)} C_2^{(2)} = {1 \over 2} (A+B) ~,
\ea
which clearly satisfy the condition
\be
{\Sigma}^2 = M^2 - A^2 + B^2 ~.
\ee
Also, using an appropriate shift of $\beta$ we may fix,
\be
\Sigma = {1 \over 2} (C_0^{(1)} - C_0^{(2)}) ~, ~~~~~
Z = {1 \over 2} (C_0^{(1)} + C_0^{(2)}) ~.
\ee

We introduce now the change of variables
\be
\alpha = \sqrt{(r - M)^2 - {\Sigma}^2} \sin \theta ~, ~~~~~
\beta - Z = (r - M) \cos \theta ~,
\ee
and substitute for ${\mu}_1$ and ${\mu}_2$. Provided that we choose
the
solutions ${\mu}_{k+}$ of (4.3), we obtain
\be
{\mu}_1 = 2(r - M + \Sigma) {\sin}^2 {\theta \over 2} ~, ~~~~~
{\mu}_2 = 2(r - M - \Sigma) {\sin}^2 {\theta \over 2} ~.
\ee
Hence, the 2-soliton solution depends only on three moduli $A$, $B$
and $M$,
while $\Sigma$ is fixed by (4.10) and $Z$ does not appear anywhere.
It is also useful to introduce the change of variable
\be
t = - \tau + 2A \varphi
\ee
and identify $t$ with the time coordinate. Then, substituting in
(4.4)--(4.8)
we may compute the explicit form of the 2-soliton metric in the
coordinates $(r, \theta, \varphi, t)$.

The special case $A=B=0$ is the simplest, since the resulting
2-soliton metric
is diagonal,
\be
ds^2 = r^2 (d {\theta}^2 + {\sin}^2 \theta d{\varphi}^2) +
{r \over r - 2M} dr^2 - {r - 2M \over r} dt^2 ~,
\ee
and coincides with the Schwartzchild metric. In the more general
situation
we obtain the Kerr metric with mass parameter $M$, rotation parameter
$A$ and
NUT parameter $B$ that describes the behaviour of the 4-dim metric at
infinity.
The result of the 2-soliton construction precisely yields the
complete Kerr metric
in Boyer-Lindquist coordinates,
\ba
ds^2 & = &
{1 \over r^2 + (B - A \cos \theta)^2} \left(
- [(r - M)^2 - {\Sigma}^2 - A^2 {\sin}^2 \theta] dt^2 +
\right.\nonumber\\
& & \left. \left[{\sin}^2 \theta (r^2 + A^2 + B^2)^2 - ((r - M)^2 -
{\Sigma}^2)
(2B \cos \theta + A {\sin}^2 \theta)^2 \right] d{\varphi}^2 + \right.
\nonumber\\
& & \left. 4\left[ B \cos \theta ((r - M)^2 - {\Sigma}^2) - A
{\sin}^2 \theta
(Mr + B^2) \right] dt d \varphi \right) + \nonumber\\
& & (r^2 + (B - A \cos \theta)^2)
\left(d{\theta}^2 + {1 \over (r - M)^2 - {\Sigma}^2} dr^2 \right) .
\ea

We note finally that the Lorentzian analogue of the self-dual
Taub-NUT
metric corresponds to the limit $\Sigma = 0 = A$ and $M = \pm i B$.
This limit,
however, is somewhat peculiar in the 2-soliton sector because all
three
quantities $N_1$, $N_2$, $D$ become zero, and the two real poles
(4.13)
coincide.

\section{Euclidean (0, 2) solitons}
\setcounter{equation}{0}
\noindent
We describe now how axionic instanton solutions of
the Euclidean string background equations can be accomodated
into the present scheme. More specifically we study the moduli
space of $(0, 2)$ 2-dim solitons and determine
the specific choice of parameters that give rise to the
worm-hole (and other related backgrounds) as solitons on flat space.
It is true that these backgrounds were originally introduced as
soliton solutions of the 10-dim heterotic string theory (see for
instance [14] and references therein), but their description
as 2-dim solitons of the reduced $\beta$-function equations
using inverse scattering methods appears to be new.

We consider the general form of the $(0, 2)$ soliton solutions
starting from 4-dim flat Euclidean space. It will be convenient for
later use to consider as seed metric
\be
ds^2 = {1 \over 2 \sqrt{{\alpha}^2 + {\beta}^2}} (d{\alpha}^2 +
d{\beta}^2)
+ (\sqrt{{\alpha}^2 + {\beta}^2} - \beta) d {\psi}^2 +
(\sqrt{{\alpha}^2 + {\beta}^2} + \beta) d {\tau}^2
\ee
with $\det g = {\alpha}^2$,
instead of the Euclidean version of (4.1). However, the explicit form
of the
metric background will be used only after completing the soliton
construction in the axion-dilaton sector. The seed $\Phi$ and $b$ are
zero,
and (5.1) is flat as can be seen using the transformation
\be
\alpha = {1 \over 2} e^{2 \rho} \sin 2 \varphi ~, ~~~~~
\beta = {1 \over 2} e^{2 \rho} \cos 2 \varphi ~,
\ee
or introducing $z = \exp(\rho + i \tau) \cos \varphi$,
$w = \exp(\rho + i \psi) \sin \varphi$ that yields
$ds^2 = dz d \bar{z} + dw d \bar{w}$.

The axion-dilaton sector has Lorentzian signature (we should replace
$b^2 + e^{-4 \Phi}$ with $b^2 - e^{-4 \Phi}$ in (2.1), since the
relevant conjugate pair of field variables is now $S_{\pm} = b \pm
e^{-2 \Phi}$
instead of $b \pm i e^{-2 \Phi}$), and therefore the $(0, 2)$ soliton
calculation is analogous to $(2, 0)$ metric solitons of stationary
axisymmetric gravity. There is a difference with the analysis of the
previous section, however, in that our normalized axion-dilaton seed
matrix
$\alpha {\lambda}_0 = {\rm diag}(\alpha, - \alpha)$ is ``isotropic",
while
$g_0 = {\rm diag}({\alpha}^2, -1)$ in (4.1). It is therefore
appropriate in
the present case to consider
\be
{\Psi}_0(l) = \sqrt{{\alpha}^2 - 2 \beta l - l^2}
\left(\begin{array}{ccc}
1 & & 0\\
  & &  \\
0 & & -1 \end{array} \right) .
\ee

Explicit calculation shows that the $(0, 2)$ axion-dilaton soliton
fields are
\be
e^{-2 \Phi} = 2 {\mu}_1 {\mu}_2 {N_1^{\prime} \over D^{\prime}} ~ ,
{}~~~~~
b = ({\mu}_1 - {\mu}_2) ({\alpha}^2 + {\mu}_1 {\mu}_2)
{N_2^{\prime} \over D^{\prime}} ~,
\ee
where
\ba
N_1^{\prime} & = & B^2 {\alpha}^2 ({\mu}_1 - {\mu}_2)^2 +
C^2 ({\alpha}^2 + {\mu}_1 {\mu}_2)^2 ~, \\
N_2^{\prime} & = & AC ({\mu}_1 + {\mu}_2) ({\alpha}^2 + {\mu}_1
{\mu}_2) +
BD ({\mu}_1 - {\mu}_2) ({\alpha}^2 - {\mu}_1 {\mu}_2) ~, \\
D^{\prime} & = & C^2 ({\mu}_1^2 + {\mu}_2^2)({\alpha}^2 + {\mu}_1
{\mu}_2)^2
- B^2 ({\mu}_1 - {\mu}_2)^2({\alpha}^4 + {\mu}_1^2 {\mu}_2^2) -
({\mu}_1 - {\mu}_2) \cdot \nonumber\\
& & ({\alpha}^2 + {\mu}_1 {\mu}_2) \left(AB ({\mu}_1 -
{\mu}_2)({\alpha}^2 -
{\mu}_1 {\mu}_2) + CD({\mu}_1 + {\mu}_2)({\alpha}^2 + {\mu}_1
{\mu}_2)
\right) .
\ea
Here we have used for convenience the notation
\ba
C_1^{(1)} C_1^{(2)} & = & {1 \over 2} (A + B) ~, ~~~~~
C_1^{(1)} C_2^{(2)} = -{1 \over 2} (C + D) ~,\nonumber\\
C_2^{(1)} C_2^{(2)} & = & {1 \over 2} (A - B) ~, ~~~~~
C_1^{(2)} C_2^{(1)} = {1 \over 2} (C - D) ~,
\ea
which satisfy the relation
\be
C^2 = D^2 - A^2 + B^2 ~.
\ee
So, the soliton moduli depends on four parameters in this case, $A$,
$B$, $C$
and one of the $C_0^{(k)}$, since the other one can be absorbed by
shifting
$\beta$.

Searching for an axionic instanton solution in this moduli space we
notice
that if
\be
A = \pm D ~ ; ~~~~~~ B^2 = C^2
\ee
we obtain
\be
e^{-2 \Phi} \mp b = 1 ~.
\ee
One can also prove relatively easy that there are no other axionic
instanton
solutions in this sector. Moreover, from all soliton constructions we
have
considered so far, only this example exhibits axionic instanton
solutions.

Next, we determine the explicit form of the solution choosing $A = D$
and
$B = C$ for concreteness. Substituting in (5.4) we obtain
\be
e^{2 \Phi} = {D^{\prime} \over 2 {\mu}_1 {\mu}_2 N_1^{\prime}} =
1 - {A \over B} {({\mu}_1 - {\mu}_2)({\alpha}^2 + {\mu}_1 {\mu}_2)
\over {\mu}_2 ({\alpha}^2 + {\mu}_1^2)} ~.
\ee
Further manipulation using (4.3) yields
\be
e^{2 \Phi} = 1 - 2 {A \over B} (C_0^{(1)} - C_0^{(2)})
{1 \over {\mu}_1 + {{\alpha}^2 \over {\mu}_1}}
\ee
and so the dilaton depends only on ${\mu}_1$. Choosing the solution
${\mu}_+$ and taking into account that ${\mu}_- = -
{\alpha}^2/{\mu}_+$
in this case, we finally arrive at the result
\be
e^{2 \Phi} = 1 + {M \over \sqrt{(C_0^{(1)} - \beta)^2 + {\alpha}^2}}
{}~ ;
{}~~~~~ M = (C_0^{(2)} - C_0^{(1)}) {A \over B} ~.
\ee
$C_0^{(1)}$ can be absorbed by shifting $\beta$, and hence for all
purposes
it may be set equal to zero.

We also note for completeness that for $A=D$, but $B=-C$, the dilaton
depends only on ${\mu}_2$ and the result turns out to be essentially
the same up
to an interchange of $C_0^{(1)}$ and $C_0^{(2)}$.
Similar remarks apply to the axionic anti-instanton case $A = -D$
with
$B = \pm C$.

The solution we have obtained in this fashion can be put together
with
the flat space metric (5.1) to yield in the Einstein frame the 4-dim
string background
\ba
ds^2 & = & e^{2 \rho} (d{\rho}^2 + d{\varphi}^2 + {\sin}^2 \varphi
d{\psi}^2 + {\cos}^2 \varphi d{\tau}^2) ~,\\
e^{2 \Phi} & = & 1 + 2M e^{-2 \rho} ~, ~~~~~ b = e^{-2 \Phi} + {\rm
const.}
\ea
We have used the change of coordinates (5.2) and the fact that for
axionic instantons there is no contribution to the conformal factor
$f$ coming from the $\lambda$
sector (see eqs. (2.8), (2.9)). This
configuration is actually known as the worm-hole solution. A related
version of it, which represents only the throat of the worm-hole and
coincides
with the semi-classical geometry of the $SU(2) \times U(1)$ WZW
model,
consists of a dilaton field with different boundary condition in that
the constant term in (5.16) is missing, but with the same
Einstein metric. These two
models are related to each other by an $SL(2, R)$ transformation (the
continuous counterpart of S-duality), which keeps the axionic
instanton
condition invariant, say $S_-$ constant, and simply shifts $e^{2
\Phi}$
by a constant.

We finally point out that higher solitonic excitations of the
axion-dilaton sector might be interesting to consider in this case.
We have determined the solution of
the linearized system
for the $\lambda$ sector of the semi-classical background
of the model $SU(2) \times U(1)$,
\be
{\Psi}_{{\rm wh}} (l) = \left(\begin{array}{ccc}
{{\alpha}^2 - \beta l \over 2
\sqrt{({\alpha}^2 + {\beta}^2)
({\alpha}^2 -2 \beta l - l^2)}} & &
\sqrt{{\alpha}^2 -2 \beta l - l^2} \\
  &  &  \\
\sqrt{{\alpha}^2 -2 \beta l - l^2} & & 0 \end{array} \right) .
\ee
However, we are not in a position at this moment to give a good
space-time
interpretation to its multi-soliton excitations.
It might turn out an interesting
geometrical problem and we hope to return to it
elsewhere.

\section{Conclusions and discussion}
\setcounter{equation}{0}
\noindent
In this paper we have considered solitonic solutions of the
2-dim reduced $\beta$-function equations for gravitational
string backgrounds with axion and dilaton fields. We found
that many known solutions that admit an exact conformal
field theory description arise as simple solitonic
excitations of flat space, or its T-dual face, depending
on the particular examples. These backgrounds include
cosmological solutions, as well as 4-dim black-holes and
worm-holes. It should be also interesting to consider
further generalizations to colliding gravitational plane
wave solutions of string theory, thus extending the
results of [20] to strings.

A key point in implementing our construction is the
non-trivial interplay between the coordinate and the
solution generating (Geroch-type) transformations.
These two kind of transformations do not commute with
each other, and we should always use a coordinate
system in the seed background with non-constant
$\det g$. The Geroch transformations leave $\det g$
invariant, and only after the solitonic dressing
is performed we may
transform the new background into any other suitable
coordinate system. Note that the action of the Geroch
group would be trivial if we were starting from flat
space in Cartesian coordinates because there are no
other physical solutions of the Ernst $\sigma$-model
with $\det g = \pm 1$ that are also compatible with
the equations for the conformal factor $f$. Actually,
the conformal factor has a very special role
incorporating the effects of 2-dim gravity in the
reduced form of the string background equations.

For certain discrete values of the soliton moduli the
underlying string backgrounds are equivalent as exact
conformal field theories, as it was demonstrated
explicitly for the general one-parameter form of
the Nappi-Witten universe. This happens because the
corresponding $O(2, 2; Z)$ T-duality transformations
[17] act on the soliton moduli as
described. Since the soliton transformations
provide specific elements of the string Geroch group,
i.e. the current group $\hat{O} (2,2)$ in the
simplest case under discussion,
and they also contain S transformations and their
intertwining with T,
our description suggests that other discrete
remnants of this infinite dimensional group could
act as U-dualities in string compactifications to
two dimensions [4, 10].

Further progress in this direction certainly requires
extending the formalism to include gauge fields as
well. For example, it will be interesting to describe
4-dim black-holes with electric charge $Q$ as 2-dim
solitons of an Ernst-type $SU(2, 1)$ $\sigma$-model.
The points in the soliton moduli space that
describe extremal black-holes are special in that
a solitonic interpretation of the resulting
configuration exists in four as well as in two
(reduced) dimensions. This should be ultimately
extended to $O(8, 24)$ $\sigma$-models, which are
applicable to heterotic string compactifications
to two dimensions [10, 11]. One should also
try to find
in this context the necessary conditions on the
Killing isometries so that the solitonic nature
of a given configuration is preserved or
attained under reduction. It seems that the
relevant distinction in the examples we have
considered so far is provided by the
translational versus the rotational character
of the corresponding Killing vector fields
(using the same terminology as in [3]).
For Euclidean black-holes, for example,
$\partial / \partial t$ is rotational unless
$M=B$ (self-dual Taub-NUT limit with $A=0$ in
the vacuum case) or $M=Q$ (extremal non-rotating
solution of the electrovacuum equations), in
which cases we find that
$\partial / \partial t$ becomes
translational. These two examples demonstrate
clearly that the reduction with respect to
rotational isometries can give rise to 2-dim
solitonic configurations, in the sense
described above, even if the 4-dim configurations
are not so. A characteristic feature of
rotational isometries in supersymmetric
backgrounds is that the fermions
depend on the Killing coordinates, while the
bosonic fields do not, and hence it is like
having a coordinate dependent compactification.

The soliton solutions of the 4-dim theory (BPS
states) are quantum mechanically stable having
manifest space-time supersymmetry. On the other
hand, the 2-dim solitonic interpretation of a
given configuration was only used here in the
context of the inverse scattering method
with no reference to supersymmetry. Hence, it
arises a natural question to find the special
properties of these 2-dim solitons with
respect to 2-dim reductions of the
space-time supersymmetry algebra. To the best
of our knowledge the only helpful
results that exist
in the literature are contained in [22], where
the 2-dim reduction of maximal supergravity
(and some of its consistent truncations) were
described in terms of Lax pairs and integrable
structures. We briefly mention that the local
supersymmetry transformations can be bosonized
in two dimensions and they can be included
as Kac-Moody variations into the corresponding
infinite dimensional hidden symmetry
(Geroch-type)
transformations. Also, the variation of the
conformal factor that accounts for the 2-dim
gravitational effects in this case can be
included systematically using central extensions
of the associated current groups [23]. Further
work is certainly required in this direction,
putting together in a more constructive way
the vast variety of
soliton solutions obtained by the inverse
scattering methods with the supersymmetric
configurations of the 2-dim reduced
supergravity. It is conceivable that 11-dim
supergravity will be singled out in this
line of thought, according to earlier
expectations by Nicolai [22].

Summarizing, the 2-dim reduced sector of string
theory is quite rich in symmetry due to the
integrability structure of the lowest order
effective theory. Better understanding of its
soliton solutions are worthy 
in the light of the recent
developments in non-perturbative string
theory, in order to establish and
explore the meaning of infinitely many
U-dualities in the spectrum. This particular
sector of string theory, where many more
connections are expected to exist than in other
sectors, could also help in the long run to
expose better the right structures that are
needed
for the ultimate formulation of string theory.
Could it be an exactly solvable sector? We
hope to return to it elsewhere.  

\vskip2cm

\centerline{\bf REFERENCES}
\begin{enumerate}
\item C. Hull and P. Townsend, Nucl. Phys. \underline{B438}
(1995) 109.
\item E. Witten, Nucl. Phys. \underline{B443} (1995) 85.
\item I. Bakas, Phys. Lett. \underline{B343} (1995) 103.
\item I. Bakas, Nucl. Phys. \underline{B428} (1994) 374.
\item R. Geroch, J. Math. Phys. \underline{12} (1971) 918;
ibid \underline{13} (1972) 394.
\item E. Cremmer and B. Julia, Nucl. Phys. \underline{B159}
(1979) 141; B. Julia, in ``Superspace and Supergravity",
eds S. Hawking and M. Rocek, Cambridge University Press,
Cambridge, 1981.
\item N. Marcus and J. Schwarz, Nucl. Phys. \underline{B228}
(1983) 145.
\item H. Nicolai, in ``Proceedings of 30th Schladming Winter
School: Recent Aspects of Quantum Fields", DESY preprint
91-038.
\item A. Sen, Nucl. Phys. \underline{B434} (1995) 179.
\item A. Sen, Nucl. Phys. \underline{B447} (1995) 62.
\item J. Schwarz, Nucl. Phys. \underline{B454} (1995) 427.
\item C. Nappi and E. Witten, Phys. Lett. \underline{B293}
(1992) 309.
\item C. Kounnas and D. Lust, Phys. Lett. \underline{B289}
(1992) 56.
\item C. Callan, J. Harvey and A. Strominger, Nucl. Phys.
\underline{B359} (1991) 611.
\item C. Kounnas, Phys. Lett. \underline{B321} (1994) 26.
\item I. Antoniadis, S. Ferrara and C. Kounnas, Nucl. Phys.
\underline{B421} (1994) 343.
\item A. Giveon and A. Pasquinucci, Phys. Lett. \underline{B294}
(1992) 162.
\item M. Rocek, K. Schoutens and A. Sevrin, Phys. Lett.
\underline{B265} (1991) 303.
\item V. Belinski and V. Sakharov, Sov. Phys. JETP \underline{48}
(1978) 985; ibid \underline{50} (1979) 1.
\item V. Ferrari, J. Ibanez and M. Bruni,
Phys. Rev. \underline{D36} (1987) 1053; V. Ferrari and J. Ibanez,
Gen. Rel. Grav. \underline{19} (1987) 405.
\item C. Cosgrove, J. Math. Phys. \underline{21} (1980) 2417.
\item H. Nicolai, Phys. Lett. \underline{B194} (1987) 402; H. Nicolai
and N. Warner, Comm. Math. Phys. \underline{125} (1989) 369.
\item P. Breitenlohner and D. Maison, Ann. Inst. Poincare
\underline{46} (1987) 215.
\end{enumerate}
\end{document}